# Velocity profiles of Osipkov–Merritt models

C. Marcella Carollo,[1] P. Tim de Zeeuw[1] and Roeland P. van der Marel[2,3]

[1] *Sterrewacht Leiden, Postbus 9513, 2300 RA Leiden, The Netherlands*
[2] *Institute for Advanced Study, Olden Lane, Princeton, NJ 08540, USA*
[3] *Hubble Fellow*

**ABSTRACT**
A simple algorithm is presented for the calculation of the projected line-of-sight velocity profiles (VPs) of non-rotating anisotropic spherical dynamical models with a phase-space distribution function of the Osipkov–Merritt type. The velocity distribution in these models is isotropic inside the anisotropy radius $r_a$ and becomes increasingly radially biased at larger radii. VP shape parameters are presented for a family of models in which the luminous mass density has a power-law cusp $\rho_L \propto r^{-\gamma}$ at small radii and a power-law fall-off $\rho_L \propto r^{-4}$ at large radii. Self-consistent models and models in which the luminous matter is embedded in a dark halo are discussed. The effects of changes in the cusp slope $\gamma$ and in the anisotropy radius $r_a$ are documented, and the area in the $(\gamma, r_a)$-plane that contains physical models is delineated. The shapes of the VPs of the models show a considerable (and observable) variation with projected galactocentric radius. These models will be useful for interpreting the data on the VP shapes of elliptical galaxies that are now becoming available.

## 1 INTRODUCTION

The stellar kinematics of elliptical galaxies provides information on the process of galaxy formation and on the presence and properties of dark matter components, such as a dark halo or a central black hole. Traditionally, only the mean stellar line-of-sight velocity and velocity dispersion could be determined from observed galaxy spectra. Only recently (e.g., Franx & Illingworth 1988) has it become possible to derive from observations the full projected line-of-sight velocity distribution, henceforth referred to as the *velocity profile* (VP). A variety of data analysis techniques is now available for this purpose (Bender 1990; Rix & White 1992; van der Marel & Franx 1993; Kuijken & Merrifield 1993; Saha & Williams 1994). These have been used to determine VP shapes for a significant number of galaxies (e.g., Bender, Saglia & Gerhard 1994; van der Marel et al. 1994a; Carollo et al. 1995). Stellar dynamical models are required to interpret such observations.

Combined photometric and kinematic data generally do not uniquely constrain the stellar phase–space distribution function (DF) $f$ (e.g., Dejonghe 1987). The conventional approach is therefore to calculate observable properties for special classes of DF's, and compare these to data. In this paper we restrict ourselves to non-rotating spherical systems. Therefore the DF depends on the energy and squared angular momentum per unit mass, $f = f(E, L^2)$ (Binney & Tremaine 1987, hereafter BT). Many DFs of this type have been discussed in the literature, some chosen for mathematical convenience, others for physical plausibility (e.g., Michie 1963; Osipkov 1979, hereafter O79; Merritt 1985a,b, hereafter M85; Stiavelli & Bertin 1985; Cuddeford 1991; Gerhard 1991; Louis 1993).

One constraint to be satisfied by any realistic DF is that it matches the observed surface brightness profiles of elliptical galaxies. De Vaucouleurs' (1948) $R^{1/4}$ law generally provides a remarkably good fit to ground–based observations. However, the corresponding mass density and gravitational potential cannot be evaluated analytically, which complicates the construction of dynamical models. For this reason, various simple analytical potential–density pairs have been proposed that in projection approximate the $R^{1/4}$ law. Many of the popular models have a constant mass density core. Jaffe (1983) and Hernquist (1990, hereafter H90) realized that such models do not fit the $R^{1/4}$ law at small radii, and proposed simple analytical potential–density pairs with a central mass density cusp. In their models the mass density inside the half-mass radius varies as $\rho \propto r^{-2}$ and $\rho \propto r^{-1}$, respectively. Recently, observations with the Hubble Space Telescope have revealed that essentially all early-type galaxies have central surface brightness cusps, with a large variety of cusp slopes (e.g., Lauer et al. 1992; Crane et al. 1993; Stiavelli, Møller & Zeilinger 1993). This independently prompted Dehnen (1993, hereafter D93), Tremaine et al. (1994, hereafter T94) and Carollo (1993; hereafter C93), to study generalizations of the Jaffe and Hernquist models with $\rho \propto r^{-\gamma}$ at small radii. These 'γ-models' have a luminous mass density of the form:

$$\rho_L(r) = \frac{(3-\gamma)aM_L}{4\pi r^\gamma (r+a)^{4-\gamma}}, \tag{1}$$

with $a$ a scale-length, $M_L$ the total luminous mass of the system and $0 \leq \gamma < 3$. The profile falls off proportional to $r^{-4}$ at large radii. The models of Jaffe and Hernquist are recovered for $\gamma = 2$ and $\gamma = 1$, respectively. The gravitational potential associated with the luminous density (1) is obtained from the Poisson equation, and has the simple form (C93; D93; T94):

$$\Phi_L(r) = \begin{cases} \dfrac{GM_L}{a}\ln\dfrac{r}{r+a}, & \text{for } \gamma = 2, \\ \dfrac{GM_L}{(2-\gamma)a}\left[\left(\dfrac{r}{r+a}\right)^{2-\gamma} - 1\right], & \text{for } \gamma \neq 2, \end{cases} \tag{2}$$

where $G$ is the gravitational constant.

Many dynamical properties of the γ-models with an isotropic DF, $f = f(E)$, were given by D93 and T94. Here we discuss the dynamical properties, and particularly the VPs, of γ-models with DFs of the anisotropic Osipkov-Merritt (O79; M85) type. Our study is motivated by the fact that only few detailed VP studies for anisotropic systems are available in the literature (e.g., Dejonghe 1987; Dehnen & Gerhard 1993; Gerhard 1993; van der Marel & Franx 1993).

In Section 2 we show that the VPs of models with an Osipkov-Merritt DF can be written as a one-dimensional integral. This integral is no more complicated than the corresponding integral for the isotropic case (included in the Osipkov-Merritt models as a limiting case). In Section 3 we discuss the range of anisotropy radii for which the Osipkov-Merritt γ-models are physical and likely to be stable, and we present the VPs for models with and without dark halos. To facilitate the comparison with observational data we characterize the predicted VP shapes by means of their Gauss-Hermite moments (e.g., van der Marel & Franx 1993). We study the variation of the Gauss-Hermite moments as a function of projected radius, cusp steepness γ and (radial) anisotropy of the model. Concluding remarks follow in Section 4.

Most of the results presented here are contained in C93. While work on the present manuscript was in progress, two papers appeared by Hiotelis (1994a,b). He studied some properties of Osipkov-Merritt DFs for models with the Jaffe luminosity density ($\gamma = 2$), both with and without a dark halo. He briefly discussed the VPs of the models, calculated by 'brute force' numerical evaluation of three-dimensional integrals. Our models are more general than his, in that we consider a range of values for $\gamma$ and discuss in more detail the dependence of the observables on the model parameters.

## 2 OSIPKOV–MERRITT MODELS

We derive a general expression for the VPs of Osipkov-Merritt models, valid for an arbitrary luminous mass density $\rho_L(r)$, and for both self-consistent and non self-consistent models. In the former case only the luminous mass density contributes to the gravitational potential: $\Phi(r) = \Phi_L(r)$. In the latter case there is an additional contribution from a dark matter component: $\Phi(r) = \Phi_L(r) + \Phi_D(r)$. Following BT we define the relative potential of the system as $\Psi(r) \equiv -\Phi(r)$. The relative potential at infinity, $\Psi_\infty$, can either have a finite value (such as in a system of finite mass) or be $-\infty$ (such as in a system embedded in a logarithmic potential). In the former case we set $\Psi_\infty$ to zero by adding a suitable constant of integration, without loss of generality.

We employ two coordinate systems for our models. Projected dynamical quantities are described with cylindrical coordinates $(R, \phi, z)$, with $R$ the projected radius on the sky, $\phi$ the azimuthal angle on the sky and $z$ along the line of sight. Internal dynamical quantities are described with the corresponding spherical coordinates $(r, \theta, \phi)$.

### 2.1 Distribution functions

We consider the special class of non-rotating models for which the DF is of the Osipkov-Merritt type (O79; M85; BT):

$$f(E, L^2) = f(Q), \quad \text{with} \quad Q \equiv E - (L^2/2r_a^2), \tag{3}$$

and the additional constraint that

$$f(Q) = 0 \quad \text{for} \quad Q \leq \Psi_\infty. \tag{4}$$

Here $E = \Psi - (v^2/2)$ is the 'relative energy' per unit mass and $L^2 = r^2(v_\theta^2 + v_\phi^2)$ is the squared angular momentum per unit mass. In a system with finite total mass, $E$ is the binding energy of a star per unit mass, and the constraint (4) implies that there are no unbound stars in the system and no stars on certain high angular momentum bound orbits. The velocity dispersion anisotropy as measured by Binney's (1980) β-parameter is (M85):

$$\beta(r) \equiv 1 - \frac{\langle v_\theta^2 \rangle + \langle v_\phi^2 \rangle}{2\langle v_r^2 \rangle} = \frac{r^2}{r_a^2 + r^2}. \tag{5}$$

The free parameter $r_a$ in the definition of $Q$ is therefore called the anisotropy radius. Inside $r_a$ the velocity distribution is nearly isotropic, while outside $r_a$ it becomes increasingly more radially anisotropic. For fixed $\rho_L$, the unique isotropic model with $f = f(E)$ is recovered in the limit $r_a \to \infty$.

The DF for the Osipkov-Merritt models follows from the luminous mass density and the potential through an Abel transform, similar to the one in Eddington's formula for $f(E)$ in the isotropic case (BT):

$$f(Q) = \frac{1}{2\pi^2\sqrt{2}} \frac{dG(Q)}{dQ}, \tag{6}$$

where the function $G(Q)$ is defined as:

$$G(Q) \equiv \int_{\Psi_\infty}^{Q} \frac{d\rho_Q}{d\Psi} \frac{d\Psi}{\sqrt{Q-\Psi}}, \tag{7}$$

and we have written

$$\rho_Q(r) \equiv \left(1 + \frac{r^2}{r_a^2}\right)\rho_L(r). \tag{8}$$

For the self-consistent γ-models with $\rho_L$ as in equation (1), $G(Q)$ is elementary for special values of $\gamma$ (H90; D93; T94). In the general case, $G(Q)$ is most easily computed using the radius rather than the potential as the independent variable, i.e.:

$$G(Q) = -\int_{r(Q)}^{\infty} \frac{d\rho_Q}{dr} \frac{dr}{\sqrt{Q-\Psi(r)}}. \tag{9}$$

The inversion $r = r(\Psi)$ required to obtain the lower bound of the integral must generally be evaluated numerically. It can be evaluated analytically for the self-consistent $\gamma$-models with potential (2):

$$r = \begin{cases} a \big/ \left[\exp\left(\frac{a\Psi}{GM_L}\right) - 1\right], & (\gamma = 2), \\ a \big/ \left\{\left[1 - \frac{a\Psi(2-\gamma)}{GM_L}\right]^{1/(\gamma-2)} - 1\right\}, & (\gamma \neq 2). \end{cases} \quad (10)$$

The DF $f(Q)$ given by equation (6) can be evaluated for any luminous mass density $\rho_L(r)$, potential $\Psi(r)$ and anisotropy radius $r_a$. It is physical when the function $G$ is a monotonically increasing function of $Q$.

## 2.2 Velocity dispersions and total kinetic energy

The velocity dispersions $\langle v_r^2 \rangle$ and $\langle v_\theta^2 \rangle = \langle v_\phi^2 \rangle \equiv \frac{1}{2}\langle v_\perp^2 \rangle$ as a function of radius in Osipkov–Merritt models can be computed by integration of $v_r^2 f(Q)$ and $v_\theta^2 f(Q)$, respectively, over all velocities. Alternatively, they can be found by integration of the single Jeans equation for spherical models with $f = f(E, L^2)$, given in eq. (4–30) of BT. The result is (cf. M85):

$$\rho_L \langle v_r^2 \rangle = -\frac{r_a^2}{r_a^2 + r^2} \int_r^\infty dr \, \rho_Q(r) \frac{d\Psi(r)}{dr}, \quad (11)$$

$$\rho_L \langle v_\perp^2 \rangle = \frac{2 r_a^2}{r_a^2 + r^2} \rho_L \langle v_r^2 \rangle.$$

Calculation of $\rho_L \langle v_r^2 \rangle$ can be done conveniently by transforming to $r = r(\Psi)$. For the self-consistent $\gamma$-models the integral for $\rho_L \langle v_r^2 \rangle$ is elementary when $4\gamma$ is an integer.

The total kinetic energies $T_r$ and $T_\perp$ in the radial and tangential directions, respectively, are defined as

$$T_r = 2\pi \int_0^\infty dr \, r^2 \rho_L \langle v_r^2 \rangle, \qquad T_\perp = 2\pi \int_0^\infty dr \, r^2 \rho_L \langle v_\perp^2 \rangle. \quad (12)$$

The ratio $2T_r/T_\perp$ equals 1 in an isotropic model, and is larger than one in $f(Q)$ models with finite $r_a$. Its value is a global measure of the dominance of radial over tangential motion. Models with too much radial motion are likely to be unstable to triaxial deformation by means of the radial orbit instability (Polyachenko 1981; Fridman & Polyachenko 1984). A useful rule of thumb (although not an exact instability criterion) is that $2T_r/T_\perp$ should not exceed some (model dependent) threshold value (e.g., Barnes 1985; Merritt & Aguilar 1985; Barnes, Goodman & Hut 1986; Merritt 1987; Aguilar 1988).

Substitution of expressions (11) in the definitions (12) and exchange of the order of integration leads to

$$T_r = -2\pi r_a^2 \int_0^\infty dr \left[r - r_a \arctan \frac{r}{r_a}\right] \rho_Q(r) \frac{d\Psi(r)}{dr},$$

$$T_\perp = -\frac{W}{2} - T_r, \quad (13)$$

where $W$ is the total potential energy

$$W = 4\pi \int_0^\infty dr \, r^3 \rho_L(r) \frac{d\Psi(r)}{dr}. \quad (14)$$

It follows that $T_r + T_\perp = -W/2$, which is the virial theorem. For a self-consistent model $d\Psi(r)/dr = -GM(r)/r^2$, where $M(r) = 4\pi \int_0^r dr \, r^2 \rho_L(r)$ is the (luminous) mass enclosed inside radius $r$. Both $M(r)$ and $W$ can be given explicitly for the $\gamma$-models (D93; T94):

$$M(r) = \frac{M_L r^{3-\gamma}}{(r+a)^{3-\gamma}}, \quad (15)$$

and

$$W = -\frac{GM_L^2}{2a(5-2\gamma)}, \quad \text{for} \quad \gamma < 5/2. \quad (16)$$

$W$ is infinite when $\gamma \geq 5/2$. We study the behaviour of $2T_r/T_\perp$ for the Osipkov–Merritt $\gamma$-models in Section 3.1.

## 2.3 Evaluation of the velocity profiles

By definition, the velocity profile VP is obtained by integrating the DF $f(E, L^2)$ over the line of sight and over the two velocity components that lie in the plane of the sky. Hence, the normalized VP is

$$\mathrm{VP}(R, v_z) \equiv \frac{1}{\Sigma(R)} \int dz \iint dv_R \, dv_\phi \, f(Q), \quad (17)$$

where $\Sigma$ is the projected surface density:

$$\Sigma(R) \equiv \int dz \iiint dv_R \, dv_\phi \, dv_z \, f(Q) = 2 \int_R^\infty \frac{r \, dr}{\sqrt{r^2 - R^2}} \rho_L(r). \quad (18)$$

The VP does not depend on the azimuthal angle $\phi$ on the plane of the sky, because of the spherical symmetry. We now show how the triple integral (17) can be simplified.

Using $(v_R, v_\phi, v_z)$ as coordinates in velocity space, the squared angular momentum is:

$$L^2 = r^2 [(v_R \cos\theta - v_z \sin\theta)^2 + v_\phi^2]. \quad (19)$$

Hence,

$$Q = \Psi(r) - \frac{1}{2}\left[v_R^2 \left(1 + \frac{r^2}{r_a^2}\cos^2\theta\right) + v_\phi^2\left(1 + \frac{r^2}{r_a^2}\right) \right. \\ \left. + v_z^2\left(1 + \frac{r^2}{r_a^2}\sin^2\theta\right) - 2 v_R v_z \frac{r^2}{r_a^2}\cos\theta \sin\theta\right]. \quad (20)$$

At a fixed point in space, $Q$ is quadratic in the velocities, and can be cast in diagonal form by replacing $(v_R, v_\phi)$ by new velocity space coordinates $(\omega, \xi)$, defined through

$$v_R = \frac{r^2 \cos\theta \sin\theta}{r_a^2 + r^2 \cos^2\theta} v_z + \sqrt{\frac{r_a^2 + r^2}{r_a^2 + r^2 \cos^2\theta}} \, \omega \sin\xi, \quad (21)$$

$$v_\phi = \omega \cos\xi.$$

Substitution in equation (20) yields:

$$Q = \Psi(r) - \frac{1}{2}\left(1 + \frac{r^2}{r_a^2}\right)\left[\omega^2 + \frac{r_a^2}{r_a^2 + r^2 - R^2} v_z^2\right], \quad (22)$$

which does not depend on $\xi$. We therefore transform the integral over $dv_R \, dv_\phi$ in equation (17) into an integral over $d\xi \, d\omega$:

$$\iint f(Q) \, dv_R \, dv_\phi = \sqrt{\frac{r_a^2 + r^2}{r_a^2 + r^2 - R^2}} \int_0^{2\pi} d\xi \int_0^\infty \omega \, d\omega \, f(Q). \quad (23)$$

We carry out the trivial integration over $\xi$, and transform the integration over $\omega \, d\omega$ to an integration over $Q$, by means of (22). With equation (4) this gives

$$\iint f(Q) \, dv_R \, dv_\phi = 2\pi \, g(r, R) \int_{\Psi_\infty}^{Q_{\max}} f(Q) \, dQ, \qquad (24)$$

where

$$g(r, R) = \frac{r_a^2}{\sqrt{(r_a^2 + r^2)(r_a^2 + r^2 - R^2)}}. \qquad (25)$$

The upper limit $Q_{\max}$ follows from (22) upon substitution of $\omega = 0$:

$$Q_{\max}(r, R, v_z^2) = \Psi(r) - \frac{r_a^2 + r^2}{r_a^2 + r^2 - R^2} \frac{v_z^2}{2}. \qquad (26)$$

Substitution of expression (24) in equation (17) yields for the VP:

$$\text{VP}(R, v_z) = \frac{4\pi}{\Sigma} \int_R^\infty \frac{r g(r, R) \, dr}{\sqrt{r^2 - R^2}} \int_{\Psi_\infty}^{Q_{\max}} f(Q) \, dQ, \qquad (27)$$

where, as in equation (18), the integral over $dz$ is transformed to an integral over $dr$. With equations (6) and (7) this reduces to

$$\text{VP}(R, v_z) = \frac{\sqrt{2}}{\pi \Sigma} \int_{r \in I(R, v_z)} \frac{r g(r, R) \, dr}{\sqrt{r^2 - R^2}} \, G[Q_{\max}(r, R, v_z^2)], \qquad (28)$$

where $I(R, v_z)$ is the set

$$I(R, v_z) \equiv \{r : r \geq R \, \wedge \, Q_{\max}(r, R, v_z^2) \geq \Psi_\infty\}. \qquad (29)$$

In equation (28) we have used the fact that $G(\Psi_\infty) = 0$ in models with finite total luminous mass. Once $G(Q)$ has been calculated and tabulated as a function of $Q$ using equation (9), expression (28) allows the VPs to be calculated as one-dimensional integrals for all $(R, v_z)$ combinations, without the explicit evaluation of the DF. The isotropic case is recovered in the limit $r_a \to \infty$, so that $Q \to E$, $g(r, R) \to 1$ and $Q_{\max} \to \Psi(r) - v_z^2/2$.

If $\Psi_\infty = -\infty$, then $I(R, v_z)$ is the interval $[R, \infty)$ and the VP extends to infinitely large velocities. In the following we restrict ourselves to models with finite total mass, and hence $\Psi_\infty = 0$. At given $R$, the maximum absolute line-of-sight velocity value $|v_z|_{\max}(r, R)$ reachable by the stars at a given intrinsic radius $r$ is set by the requirement $Q_{\max}(r, R, v_z^2) = 0$. Hence,

$$|v_z|_{\max}(r, R) = \sqrt{\frac{2\Psi(r_a^2 + r^2 - R^2)}{r_a^2 + r^2}}. \qquad (30)$$

The $\text{VP}(R, v_z)$ extends over a finite range of velocities, $-v_t(R) \leq v_z \leq v_t(R)$, where $v_t(R)$ is the 'terminal velocity', which is obtained by maximizing $|v_z|_{\max}(r, R)$ as a function of $r$ on the interval $[R, \infty)$. At large radii, $r/R \to \infty$, one has $|v_z|_{\max}(r, R) \approx \sqrt{2\Psi(r)}$, which is a decreasing function of radius. As a result, the maximum is attained at a finite value of $R$, i.e., either at $r = R$, or at a radius $r$ for which $d|v_z|_{\max}(r, R)/dr = 0$:

$$\frac{d\Psi}{dr}(r_a^2 + r^2 - R^2) + \frac{2\Psi R^2 r}{r_a^2 + r^2} = 0. \qquad (31)$$

In an isotropic model $r_a \to \infty$, and equation (30) reduces to $|v_z|_{\max}(r, R) = \sqrt{2\Psi(r)}$. Over the interval $[R, \infty)$ this attains its maximum at $r = R$. In an isotropic model the terminal velocity is thus determined by those stars along the line of sight that are at the tangential point (deepest in the potential well), have zero energy and have their velocity vector along the line-of-sight. The angular momentum of these stars satisfies $L^2 = 2R^2\Psi(R)$. In anisotropic models with finite $r_a$ there are no such stars, because of the imposed constraint (4). It follows that the terminal velocity need not necessarily be attained by stars at the tangential point. In addition, the terminal velocity is lower than in the isotropic model.

For anisotropic models one must search (numerically) for solutions of equation (31). We found that for the $\gamma$-models there is generally either no, or only one solution. When no solutions are found, the terminal velocity is attained at $r = R$. When a solution $r_0 > R$ is found, it is attained at the radius $r = r_0$. At values of $v_z$ other than the terminal velocity, the set $I(R, v_z)$ defined by equation (29) generally reduces to an interval $[r_1, r_2]$, with $R \leq r_1 < r_2 \leq \infty$. The values of $r_1$ and $r_2$ must be calculated numerically for each combination of $(R, v_z)$.

The VPs to be discussed in the following sections were calculated by numerical integration of equation (28) for an array of $v_z$ values spaced linearly between $-v_t(R)$ and $v_t(R)$.

## 3 RESULTS FOR OSIPKOV–MERRITT $\gamma$–MODELS

First we determine which combinations of values of $\gamma$ and $r_a$ correspond to physical distribution functions $f(Q) \geq 0$, and we discuss which of these models are expected to be dynamically stable. Then we investigate the kinematic properties of the Osipkov–Merritt $\gamma$-models. We consider the self-consistent case and also investigate the effect of adding the gravitational potential of a dark halo to the potential (2). We restrict ourselves to models with $\gamma \leq 2$.

### 3.1 Existence and stability

The isotropic self-consistent $\gamma$-models ($r_a \to \infty$) all have physical distribution functions $f(E)$ (e.g., T94). When $r_a$ decreases, the fraction of stars on radial orbits in an $f(Q)$ model increases. In the limit $r_a \to 0$ only radial orbits are populated. For such a radial orbit model to be physical, the density must diverge as $r^{-2}$, or steeper, in the centre (BT). Thus, $f(Q)$ for the anisotropic $\gamma$-models with $\gamma < 2$ must become negative for sufficiently small values of $r_a$ in order to compensate for the overpopulation of the centre by the radial orbits needed to reproduce the density in the envelope of the model. We have numerically determined the area in the $(\gamma, r_a)$-plane where the Osipkov-Merritt $\gamma$-models are physical. The thick drawn solid curve in Figure 1 marks the lower boundary of the physical self-consistent models. As expected, the minimum allowed value of $r_a/a$ increases with decreasing $\gamma$, and ranges from 0 at $\gamma = 2$ to 0.44 at $\gamma = 0$.

N–body experiments have shown that $f(Q)$ models with $\gamma = 2$ are prone to the radial orbit instability when $2T_r/T_\perp \gtrsim 2.5$ or, equivalently, $r_a \lesssim 0.25a$ (Merritt & Aguilar 1985). Little is known about the stability of Osipkov-Merritt $\gamma$-models with $\gamma < 2$. Antonov's (1962) sufficient condition for stability against radial perturbations (Dejonghe & Merritt 1988) can be applied easily to the Osipkov-Merritt $\gamma$-models, but we find that it is inconclusive: it confirms that the isotropic models ($r_a \to \infty$) are stable, but gives no such guarantee for finite $r_a$.

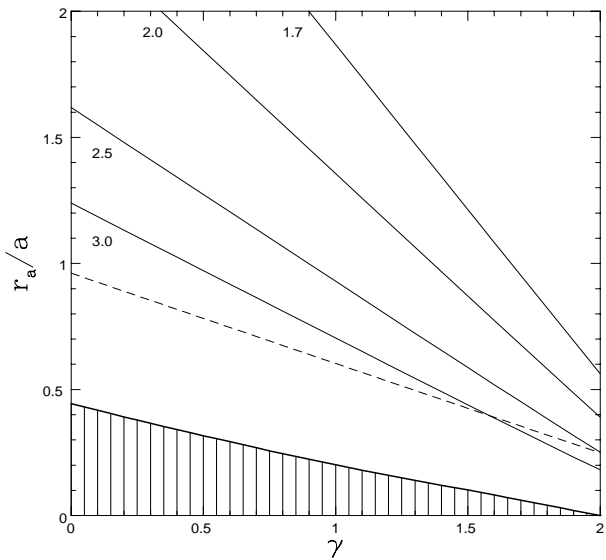

**Figure 1.** The $(\gamma, r_a)$-plane for Osipkov–Merritt $\gamma$–models. Models in the hatched area below the thick solid curve have $f(Q) < 0$ for some values of $Q$, and are unphysical. The dashed curve is $r_a = 0.25 r_h$, with $r_h$ the half-mass radius. This curve roughly represents the lower boundary of the region that corresponds to models which are not affected by the radial orbit instability. The thin solid curves are contours of constant $2T_r/T_\perp$. The contour values are indicated.

The half–mass radius $r_h$ of the $\gamma$-models is given by $r_h/a = 1/(2^{1/(3-\gamma)} - 1)$, so that $r_h = a$ when $\gamma = 2$. To good accuracy $R_{\rm eff} = 0.75 r_h$, where $R_{\rm eff}$ is the radius which contains half the projected surface density of the $\gamma$-models (D93). A naïve generalization of the stability condition $r_a \gtrsim 0.25 a$ when $\gamma = 2$ is to take $r_a = 0.25 r_h$ as the stability boundary. This relation is shown as the dashed line in Figure 1. However, the relative importance of radial orbits in models with fixed $r_a/r_h$ but different $\gamma$ increases when $\gamma$ decreases. This causes the steeper slope of the contours of constant $2T_r/T_\perp$ in Figure 1 (thin solid lines). If the stability boundary is given by $2T_r/T_\perp = 2.5$ also for $f(Q)$ models with $\gamma < 2$, then models with $r_a/a \lesssim 1.6 - 0.67\gamma$ are radial orbit unstable.

The determination of the precise location of the stability boundary in the $(\gamma, r_a)$-plane requires a normal mode analysis (e.g., Weinberg 1991), or careful N–body experiments. In what follows, we will only discuss kinematic properties of physical Osipkov-Merritt $\gamma$-models which are likely to be stable.

### 3.2 VPs for the self–consistent models

Since our models have zero mean streaming, their VPs are symmetric. In order to facilitate the comparison with observational data, we express the VP as an orthogonal Gauss–Hermite series (e.g., van der Marel & Franx 1993; Gerhard 1993) with parameters: (i) the line strength and the dispersion $\sigma$ of the best-fitting Gaussian VP; and (ii) the dimensionless Gauss-Hermite moments $z_4, z_6, \ldots$ that measure deviations from the best-fitting Gaussian with zero mean (van der Marel et al. 1994b). A value $z_4 > 0$ generally indicates that the VP is more centrally peaked than a Gaussian, a value $z_4 < 0$ that it is more flat-topped. To study the observable properties of our models, we consider only $\sigma$ and $z_4$ since the signal–to–noise ratio and instrumental resolution of spectroscopic data are often insufficient to measure $z_6$ and higher order coefficients. Typical uncertainties in published $z_4$ measurements for elliptical galaxies are of the order of 0.02.

Figure 2 illustrates the observable kinematic properties of self-consistent Osipkov-Merritt models with $\gamma = 1$. In this case the half-mass radius $r_h$ and the scale length $a$ are related through $r_h = 2.4142 a$, and the effective radius $R_{\rm eff} = 1.815 a$ (D93). The panels on the left show the effect of anisotropy on the radial $\sigma$ and $z_4$ profiles, while the panels on the right show the normalized VPs at $R = a$ and $R = r_a$, as a function of $v_z/\sigma$. Plotted are the models with $r_a = 0.3 r_h$ (dotted), $r_a = 0.5 r_h$ (short-dashed) $r_a = r_h$ (long–dashed) and $r_a = 2 r_h$ (dot-dashed). The solid curve has $r_a \to \infty$, and corresponds to the isotropic model. The model with $r_a = 0.3 r_h$ lies above the dashed line in Figure 1, but below the contour $2T_r/T_\perp = 2.5$, and its stability is not guaranteed. All VPs (including those for the isotropic model) differ appreciably from a Gaussian, and show a strong variation with radius. The $\sigma$ profiles of the anisotropic models lie below the isotropic curve at large radii, while they lie above it at small radii. When $R \gg r_a$ the motion is strongly radial, and only a small fraction of this is in the direction of the line of sight. When $R < r_a$, $\sigma$ is large because of the contribution of fore– and background stars at large galactocentric distances, which now have a substantial part of their motion directed towards the observer. As expected, the anisotropic models have $z_4$-profiles that are similar to the isotropic one when $R \ll r_a$. When $R \sim r_a$ the $z_4$-values are negative, because stars at radii $r$ slightly larger than $r_a$, though less in number, have a larger part of their motion along the line of sight, and hence contribute strongly to the VP at large velocities. This effect is stronger when $r_a$ becomes smaller than $a$, because then the density profile in this radial range falls off like $r^{-\gamma}$, and the typical local rms velocities are still increasing outwards ($v \propto r^{0.5}$ when $\rho \propto r^{-1}$). When $R \gg r_a$ the anisotropic $z_4$-profiles increasingly deviate from the isotropic curve towards more positive values at larger radii due to the increasing radial anisotropy. Now most of the motion is perpendicular to the line of sight, and hence most of the contribution to the VP is at small velocities, so that the VP is much more peaked than a Gaussian.

Similar results are obtained for other values of $\gamma$. In Figure 3 we illustrate the case $\gamma = 3/2$ (for which $r_h = 1.7024 a$ and $R_{\rm eff} = 1.276 a$; D93). Again the stability of the model with $r_a = 0.3 r_h$ is not guaranteed. The cusp in the density profile now has a steeper slope, so the stars in the centre contribute more strongly to the projected kinematic quantities.

Figure 4 shows the effects of varying the slope $\gamma$ of the inner density profile for the anisotropic self-consistent model with $r_a = a$. Plotted are the velocity profiles at $R = a$ and $R = 0.2 a$, and the radial $\sigma$ and $z_4$ profiles for $\gamma = 0.5$ (long-dashed; $R_{\rm eff} = 2.358 a$), $\gamma = 1.0$ (short-dashed), $\gamma = 1.5$ (dotted) and $\gamma = 2.0$ (solid; $R_{\text{eff}} = 0.744a$). The model with $\gamma = 0$ and $r_a = a$ is not shown since it is possibly unstable (cf. Figure 1). Varying $\gamma$ does not affect the dynamical structure at large radii, so the profiles coincide for $R \gtrsim r_a$. When $\gamma$ increases towards 2, i.e., for steeper central density profiles, the projected properties of the models are increasingly dominated by the stars in the central region. Here the velocity distribution of the $f(Q)$ models is nearly isotropic. For $\gamma \to 2$ (as in the singular isothermal sphere), it approaches a Gaussian. As a result, increasing $\gamma$ to 2 decreases the $z_4$ values inside $R = r_a$ towards 0. As expected, the asymptotic behaviour of $\sigma$ for $R \to 0$ is similar to that in the isotropic models (see also D93 and T94).

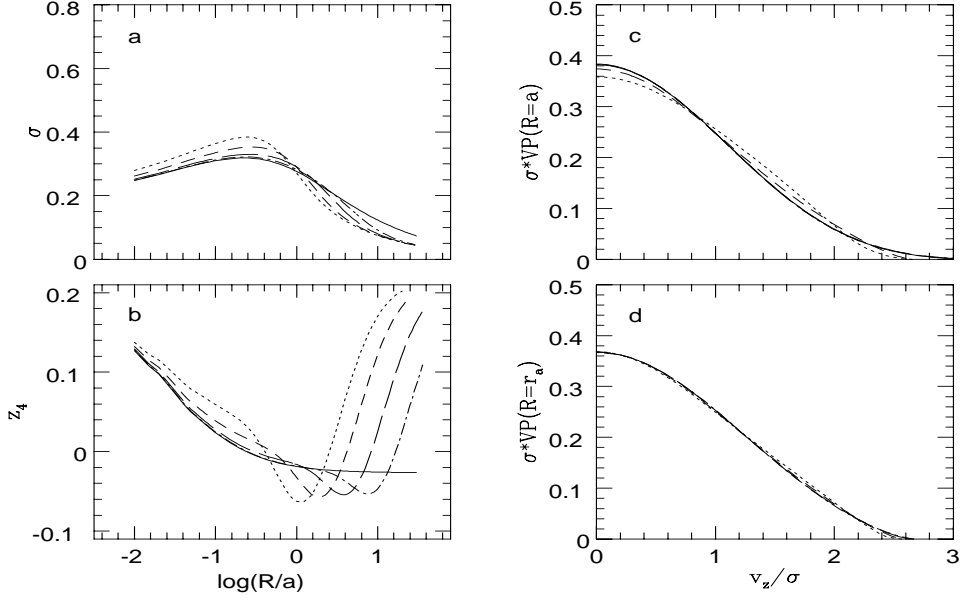

**Figure 2.** Projected kinematic profiles for the self–consistent $\gamma = 1$ (H90) model described in the text. Different curves correspond to different values for the ratio $r_a/a$ of the anisotropy radius and the length scale of the mass density: $r_a = 0.72a$ (dotted curve), $r_a = 1.21a$ (short–dashed curve), $r_a = 2.41a$ (long–dashed curve), $r_a = 4.83a$ (dot–dashed curve) and $r_a \to \infty$ (solid curve; isotropic model). These curves correspond to values of the ratio $r_a/r_h = 0.3, 0.5, 1.0, 2.0$, and $\infty$, where $r_h$ is the half–mass radius of the mass density. a) radial profiles of the projected velocity dispersion $\sigma$ in units of $\sqrt{GM_L/a}$. b) radial profiles of the fourth Gauss-Hermite coefficient $z_4$. A value $z_4 > 0$ generally indicates that the VP is more centrally peaked than a Gaussian, a value $z_4 < 0$ that it is more flat-topped. c) normalized line-of-sight velocity profiles as function of $v_z/\sigma$, at projected galactocentric distance $R = a$. d) idem, at $R = r_a$. The velocity profiles are symmetric, and only the positive velocity part is shown.

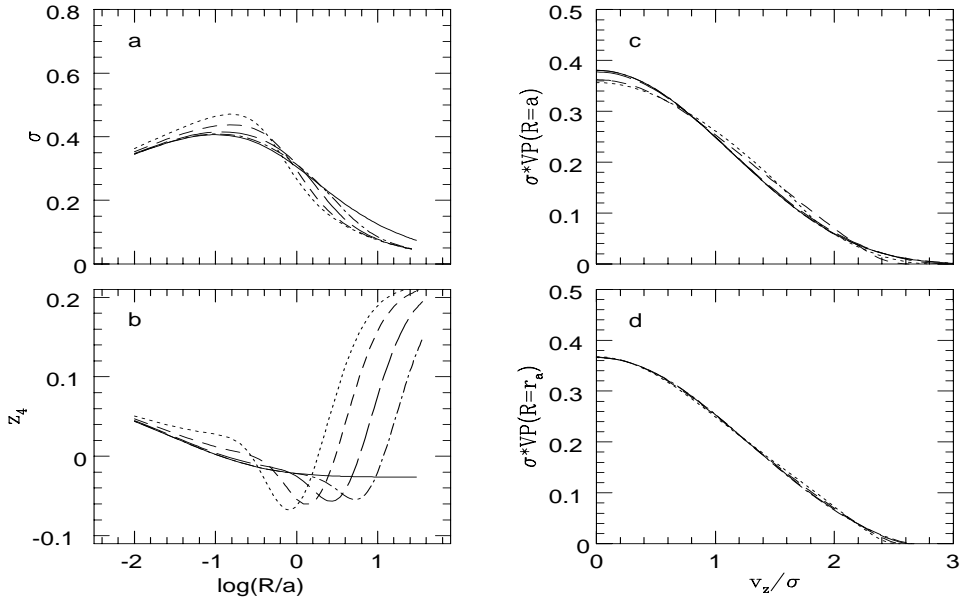

**Figure 3.** Projected kinematic profiles similar to Figure 2, but now for the self–consistent $\gamma = 3/2$ model. As in Figure 2, the curves are for the values 0.3, 0.5, 1.0, 2.0, and $\infty$ of the ratio $r_a/r_h$. These now correspond to $r_a = 0.51a, 0.85a, 1.70a, 3.40a$ and $\infty$.

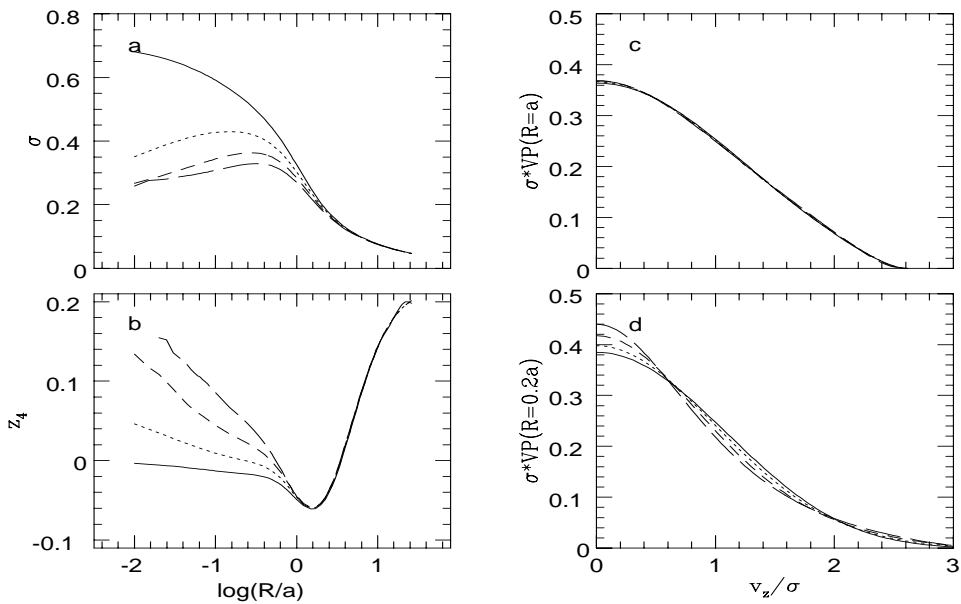

**Figure 4.** Projected kinematic profiles for self–consistent $\gamma$–models with anisotropy radius $r_a = a$. The different curves correspond to different values for the cusp steepness $\gamma$: $\gamma = 0.5$ (long–dashed curve), $\gamma = 1.0$ (short–dashed curve), $\gamma = 1.5$ (dotted curve) and $\gamma = 2.0$ (solid curve). Panels a) and b) show the $\sigma$ and $z_4$ profiles, similar to Figure 2. Panels c) and d) show the velocity profiles at projected galactocentric distances $R = a$ and $R = 0.2a$, respectively.

### 3.3 VPs for models with dark halos

We now consider the same models embedded in a dark halo. We assume that the dark potential is also represented by a $\gamma$–model, but with different scale–length $a_D > a$ and total mass $M_D$ than for the luminous matter. We only investigate the case where the value of $\gamma$ is the same for the luminous and the dark mass density. Our choice of dark halo potential results in a nearly flat rotation curve in the observationally accessible range when $a_D/a$ is sufficiently large, and has the advantage of giving a finite total mass.

The isotropic models with dark halos are physical for all values of $\gamma$. We have found that the location of the lower boundary in the $(\gamma, r_a)$–plane of the physical $\gamma$–models embedded in a dark halo with a range of masses $M_D/M_L$ and scale–lengths $a_D/a$ differs insignificantly from the thick solid curve in Figure 1. We expect that the location of the stability boundary for the models with dark halos also does not differ very much from the self–consistent boundary (cf. Stiavelli & Sparke 1991).

The VPs were again calculated as described in Section 2.2. The only differences from the self–consistent case are that now $\Psi = -\Phi_L - \Phi_D$, and that the inversion $r(\Psi)$ required for equation (9) must be evaluated numerically. Since the total mass of the halo is finite, we can again take $\Psi_\infty = 0$, and $G(\Psi_\infty) = 0$. This means that also for these non self–consistent models the calculation of the VP is done without an explicit evaluation of the DF.

Figure 5 illustrates the effect of adding a dark halo on the behaviour of the VPs at $R = a$ and $R = 0.2a$, and of $\sigma$ and $z_4$ as a function of radius for the isotropic models with $\gamma = 3/2$. Plotted are the models with $a_D = 5a$ and $M_D/M_L = 10.0$ (short–dashed), $M_D/M_L = 5.0$ (dotted) and $M_D/M_L = 0.0$ (solid). The major effect of the introduction of the dark halo is the added depth of the potential well, which results in increased velocity dispersion at all radii. The velocity distribution remains isotropic, and hence the $z_4$ values hardly change.

Figure 6 shows the similar results for the anisotropic models with $r_a = a$, again for $\gamma = 3/2$. Again the $\sigma$ values increase, as expected. The $z_4$ profiles of the models with dark halos lie systematically above the no dark halo curve, but the differences are not large. The dip in $z_4$ seen in the models without dark halo (Figure 3) is less pronounced. This is caused by the fact that the extended dark halo potential softens the radial variation of the typical orbital velocities so that the contribution to the VP of the stars at radii $r \gtrsim r_a$ is diminished.

Figure 7 shows the results for the same anisotropic models, but now with $M_D = 10M_L$, for different ratios of the

8   *C. M. Carollo et al.*

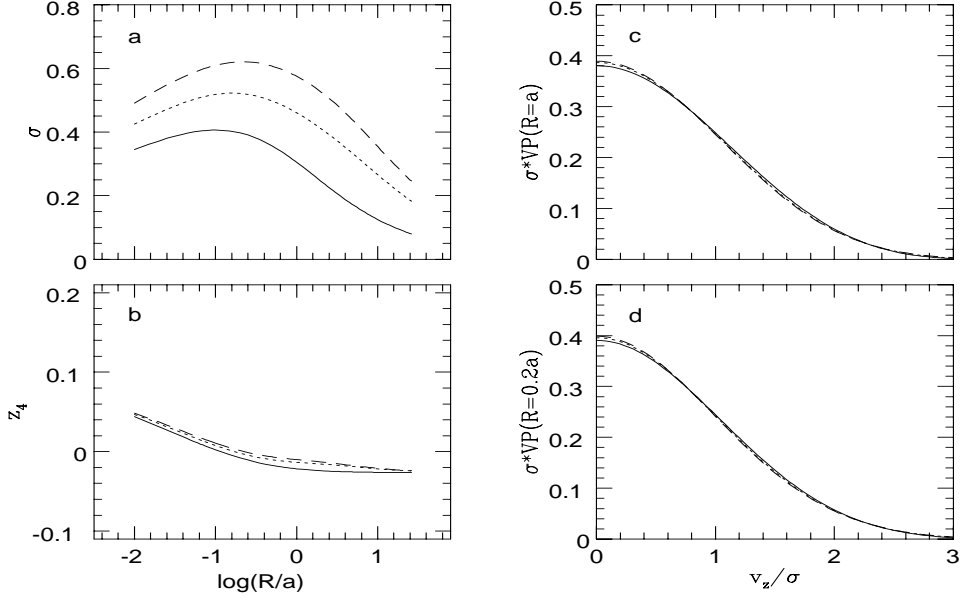

**Figure 5.** Projected kinematic profiles for isotropic models ($r_a \to \infty$) with the luminous mass density of a $\gamma = 3/2$ model, with total luminous mass $M_L$ and scale-length $a$, embedded in a dark halo that also has the mass density of a $\gamma = 3/2$ model, but has total dark mass $M_D$ and scale-length $a_D$. The different curves correspond to different values for the ratio $M_D/M_L$: $M_D/M_L = 10.0$ (short-dashed curve), $M_D/M_L = 5.0$ (dotted curve) and $M_D/M_L = 0.0$ (solid curve). All models have $a_D = 5a$. The panels a) and b) show the $\sigma$ (in units of $\sqrt{GM_L/a}$) and $z_4$ profiles, similar to Figure 2. Panels c) and d) show the velocity profiles at projected galactocentric distances $R = a$ and $R = 0.2a$, respectively.

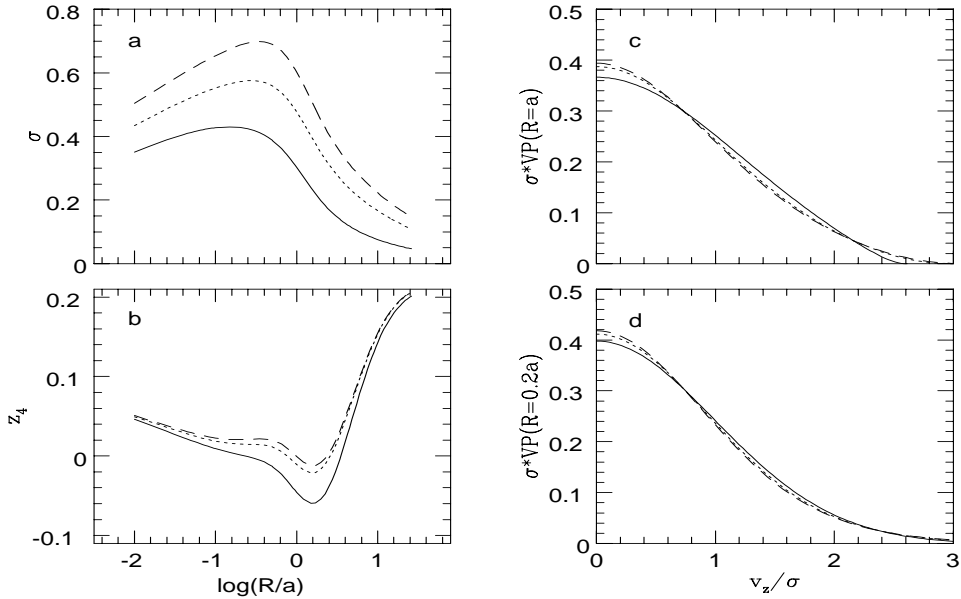

**Figure 6.** Projected kinematic profiles for $\gamma = 3/2$ models with $r_a = a$, embedded in a $\gamma = 3/2$ dark halo. The scale-length $a_D$ of the dark halo is taken equal to $5a$ in all models, while the ratios $M_D/M_L$ are varied as in Figure 5.

dark to luminous scale-lengths $a_D/a = 10, 5, 2$. The self-consistent model ($M_D = 0$) is shown for comparison (solid curve). The behaviour of the $z_4$ profiles is similar as in Figure 6, and this has the same reason: when $a_D/a$ increases, the radial variation of the potential is increasingly softened, and the dip in the $z_4$ profile is smoothed out. The $\sigma$ profiles with dark halo are all for the same total mass, and hence they converge at radii $R \gg a_D$. The differences at smaller radii are caused by the increasing central concentration of the potential when $a_D$ decreases.

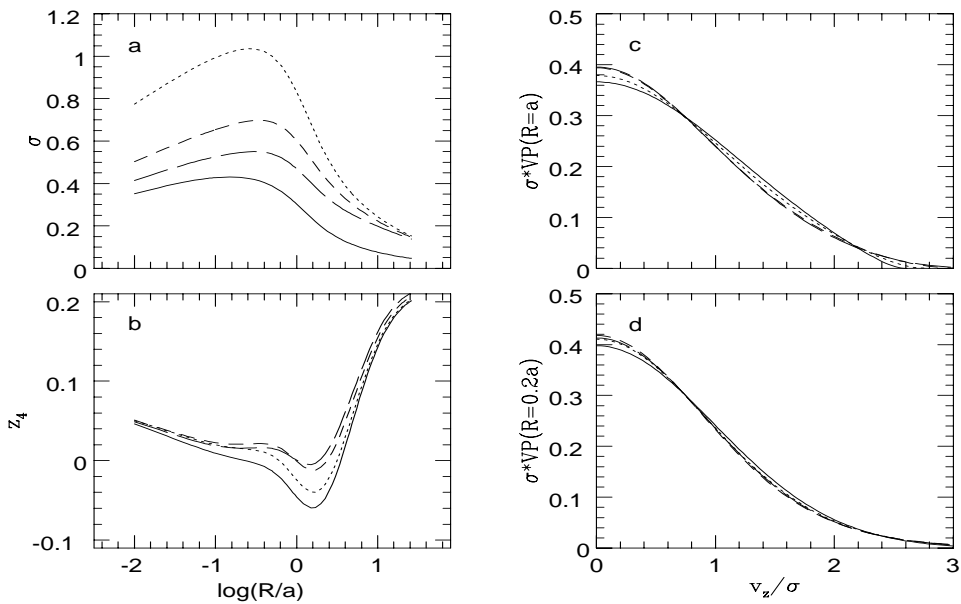

**Figure 7.** Projected kinematic profiles for $\gamma = 3/2$ models with $r_a = a$, embedded in a $\gamma = 3/2$ dark halo. This time all models have $M_D/M_L$ fixed at the value 10, and the ratio $a_D/a$ is varied. The different curves correspond to $a_D/a = 10.0$ (long-dashed curve), $a_D/a = 5.0$ (short-dashed curve), $a_D/a = 2.0$ (dotted curve). The model with $M_D = 0$ is shown for comparison (solid curve).

## 4 CONCLUDING REMARKS

The Osipkov–Merritt spherical dynamical models have the special property that their distribution functions are a function of one integral of motion: $f = f(Q)$, where $Q$ is a linear combination of the binding energy $E$ and the square of the total angular momentum, $L^2$. As such, they are the simplest spherical models with the attractive property that the intrinsic velocity distribution is isotropic in the central region, and becomes increasingly radially anisotropic going outwards. This behaviour is in good qualitative agreement with dissipationless collapse models for the formation of elliptical galaxies (e.g., van Albada 1982; Bertin & Stiavelli 1993).

We have shown that the $\gamma$-models introduced by D93, C93 and T94 correspond to physical distribution functions $f(Q) \geq 0$ in a large region of the $(\gamma, r_a)$-parameter space: only the hatched area in Figure 1 is excluded. All models with $\gamma$ and $r_a$ above the dashed line in Figure 1 are expected to be dynamically stable, but this conclusion needs to be confirmed by detailed stability analysis.

The calculation of the line–of–sight projected velocity distribution (the VP), which generally requires a triple numerical integration, reduces to a single quadrature in $f = f(Q)$ models, and therefore is no more difficult than for isotropic $f = f(E)$ models. By contrast with the isotropic models, however, the VPs of Osipkov–Merritt models generally show a strong variation with projected galactocentric radius. We have illustrated this range of VP behaviour for the $\gamma$-models. The VPs may be more peaked than a Gaussian at small radii, significantly less peaked at intermediate radii, and again more peaked at large radii. This behaviour is caused by the trade–off between the increasing radial bias at large intrinsic radii, and a decreasing contribution of the radial component of the velocity ellipsoid to the line–of–sight velocity at larger projected radii. We quantified the VP variations by the dimensionless shape–parameter $z_4$, which can vary by as much as 0.1-0.2. This is an order of magnitude larger than the typical measurement uncertainties in $z_4$. These VP variations are thus detectable with current observational techniques.

The spherical $\gamma$-models provide a good fit to the surface brightness profiles of round elliptical galaxies. We expect that the VP properties described here for $\gamma$-models with and without dark halos will be useful for the interpretation of high quality kinematic data for such galaxies.

## ACKNOWLEDGMENTS

We thank E. Emsellem and M. Stiavelli for comments on an early version of the manuscript. CMC was supported by HCM grant ERBCHBICT940967 of the European Community. RPvdM was supported by NASA through a Hubble Fellowship, #HF-1065.01-94A, awarded by the Space Telescope Science Institute which is operated by AURA, Inc., for NASA under contract NAS5-26555.

## REFERENCES

Aguilar L. A., 1988, Cel. Mech., 41, 3